\documentclass[sigconf]{acmart}

\begin{document}

\title{LLMs Aren't Human: A Critical Perspective on LLM Personality}

\author{Kim Zierahn}
\email{kim@ellisalicante.org}
\affiliation{
  \institution{ELLIS Alicante}
  \country{Spain}
}

\author{Cristina Cachero}
\email{ccachero@dlsi.ua.es}
\affiliation{
  \institution{University of Alicante}
  \country{Spain}
}

\author{Anna Korhonen}
\email{alk23@cam.ac.uk}
\affiliation{
  \institution{University of Cambridge}
  \country{United Kingdom}
}

\author{Nuria Oliver}
\email{nuria@ellisalicante.org}
\affiliation{
  \institution{ELLIS Alicante}
  \country{Spain}
}

\date{February 2026}

\setcopyright{acmlicensed}
\copyrightyear{2026}
\acmYear{2026}
\acmConference[CHI '26]{ACM CHI Conference on Human Factors in Computing Systems, Workshop on Human-Agent Collaboration}{April 13-17, 2026}{Barcelona, Spain}

\begin{abstract}
A growing body of research examines personality traits in Large Language Models (LLMs), particularly in human-agent collaboration. Prior work has frequently applied the Big Five inventory to assess LLM behavior analogous to human personality, without questioning the underlying assumptions. This paper critically evaluates whether LLM responses to personality tests satisfy six defining characteristics of personality. We find that none are fully met, indicating that such assessments do not measure a construct equivalent to human personality. We propose a research agenda for shifting from anthropomorphic trait attribution toward functional evaluations, clarifying what personality tests actually capture in LLMs and developing LLM-specific frameworks for characterizing stable, intrinsic behavior.

\end{abstract}

\keywords{Large Language Models, Human-Agent Collaboration, Anthropomorphism,  Personality, Big Five}

\maketitle

\section{Introduction}
Large Language Models (LLMs) are increasingly integrated in our daily lives, from social companionship to human-agent collaboration. To improve these interactions, researchers have endowed LLMs with personality-like characteristics \cite{huang2024revisitingreliabilitypsychologicalscales, serapiogarcia2025personalitytraitslargelanguage, huang2025designingaiagentspersonalitiespsychometric, shu2024dontneedapersonalitytest, zheng2025LMLPAlinguisticpersonalityassessment, lee2025llmsdistinctconsistentpersonality, jiang2024personallminvestigatingabilitylarge, jiang2023evaluatinginducingpersonalitypretrained, frisch2024LLMagentsininteraction}, 
suggesting that this could enhance trust, engagement, and satisfaction \cite{kuhail2024impactpersonalitycongruence, sonlu2024effectsembodimentpersonalityexpression, soderqvist2025personalitymatchedai, kovacevic2024chatbotswithattitude, moilanen2022effectofMHchatbotonuserengagement, lee2024chatfiveenhancinguserexperience}.
As LLMs show an increasing ability to mimic human behavior, scholars have used human personality tests to measure LLM behavior as a construct similar to personality in humans \cite[\emph{e.g.,}][]{serapiogarcia2025personalitytraitslargelanguage, jiang2024personallminvestigatingabilitylarge, jiang2023evaluatinginducingpersonalitypretrained, lee2025llmsdistinctconsistentpersonality, karra2023estimatingpersonalitywhitebox, huang2024revisitingreliabilitypsychologicalscales}. 
However, it remains unclear what these tests actually measure in LLMs.
Most of previous work refer to a (synthetic) ''personality'' in LLMs \cite{serapiogarcia2025personalitytraitslargelanguage, zheng2025LMLPAlinguisticpersonalityassessment, jiang2023evaluatinginducingpersonalitypretrained, hilliard2024elicitingpersonalitytraitslarge, li2024evaluatingpsychologicalsafety} or as the ability to simulate and self-report personality traits \cite{sorokovikova2024llmssimulatebigpersonality, xing2025chameleon, zou2025llmselfreportevaluatingvalidity}. A minority questions whether personality constructs meaningfully apply to LLMs at all \cite{han2025personalityillusionrevealingdissociation, romero2023llmssplitpersonality, suehr2024challengingvaliditypersonalitytests}. 
Yet there is no systematic evaluation of whether LLMs' responses to personality tests align with the definition personality.

In psychology, a person's personality denotes their "characteristic styles of thought, feeling and behavior" \cite{costa1992fivefactormodelpersonalitydisorders}. 
Personality lacks a single definition \cite{larsen2017personalitypsychology}, but many frameworks include six characteristics that define personality traits.
Firstly, personality traits are latent, \textit{internal factors}, grounded in an individual's dispositions and influenced by the environment and subjective experiences \cite{larsen2017personalitypsychology, costa1992fivefactormodelpersonalitydisorders, matthews2003personalitytraits}. They are considered relatively \textit{stable over time} \cite{larsen2017personalitypsychology, matthews2003personalitytraits} and \textit{consistent over situations} \cite{larsen2017personalitypsychology, matthews2003personalitytraits}. Personality traits describe \textit{inter-individual differences}, \emph{i.e.}, the ways in which people are different from each other \cite{costa1992fivefactormodelpersonalitydisorders, larsen2017personalitypsychology}. They influence, explain and predict \textit{behavior} \cite{larsen2017personalitypsychology, matthews2003personalitytraits}, and are appropriate \textit{descriptive summaries} of a person’s behavior \cite{larsen2017personalitypsychology, matthews2003personalitytraits}. 

The Big Five \cite{goldberg1990bigfive} is a widely used taxonomy that describes personality along five dimensions: Openness to Experience, Conscientiousness, Extraversion, Agreeableness, and Neuroticism. 
In recent years, researchers have applied this taxonomy beyond human psychology to computational social science and artificial intelligence (AI) \cite[\emph{e.g.,}][]{serapiogarcia2025personalitytraitslargelanguage, jiang2023evaluatinginducingpersonalitypretrained, salecha2024LLMsdisplaysocialdesirabilitybias, huang2025designingaiagentspersonalitiespsychometric}. 
However, this application to LLMs often assumes the framework's validity without empirical evaluation.
The literature lacks strategic analysis of whether the Big Five apply to LLMs.

This position paper addresses this gap by critically examining the anthropomorphization of LLMs from a psychological perspective.
We systematically evaluate whether LLMs satisfy the six characteristics that define personality traits in humans when responding to Big Five tests.
We propose a research agenda that clarifies what personality measurements actually capture in LLMs and shifts focus toward three distinct constructs: functional behavioral evaluation, malleable behavior relevant to human-agent interaction, and stable intrinsic characteristics of LLMs.

\section{Human Personality Frameworks Applied to LLMs: A Critical View}
This section examines whether the six defining characteristics of personality traits apply to LLMs. We examine both theoretical arguments and empirical findings from Big Five research on LLMs.

\paragraph{1. Internal Factors.}
Personality traits are understood as latent, internal factors that are deeply grounded in an individual's cognitive and affective processes. LLMs are neural networks with billions of parameters trained to predict the next probable token in a sequence of text \cite{sartori2023LMsandpsychological}. As they encode regularities and patterns from large-scale training data, one could hypothesize that LLMs develop internalized representations analogous to internal personality traits.

However, there are mixed empirical findings on consistent response patterns. Studies report acceptable internal consistency of LLMs' responses to personality tests \cite{zheng2025LMLPAlinguisticpersonalityassessment, jiang2023evaluatinginducingpersonalitypretrained, salecha2024LLMsdisplaysocialdesirabilitybias} and stable semantic relationships across different personality tests \cite{huang2025designingaiagentspersonalitiespsychometric}, suggesting that model responses reflect coherent latent patterns. Moreover, small number of studies propose that LLMs exhibit latent knowledge of personality traits \cite{molchanova2025exploringpotentiallargelanguage, bai2025scalinglawllmsimulated}.

Conversely, other work finds significant variability across test formats, challenging the notion that LLMs have internal representations of personality \cite{li2025evaluatingLLMswithpsychometrics}. Lee et al. suggest limitations in LLMs' capacities for introspection and personality self-assessment \cite{lee2025llmsdistinctconsistentpersonality}. 
Additionally, Romero et al. argue that LLMs' responses to personality tests might primarily reflect a probability distribution over words \cite{romero2023llmssplitpersonality}. Rather than expressing internal traits, LLMs replicate and aggregate behaviors observed in the training data \cite{romero2023llmssplitpersonality, tosato2025persistentinstabilityllmspersonality, karra2023estimatingpersonalitywhitebox}. 
In sum, while some studies find evidence of stable patterns in LLMs' outputs, many researchers invalidate the claim that LLMs exhibit internal factors.

\paragraph{2. Stability Over Time.}
Personality is considered stable over time, and expected to change slowly. 
Theoretical considerations and empirical evidence of comparable temporal stability in LLMs remain limited and inconclusive. 
While repeated measurements of Big Five tests on LLMs produced similar scores across time \cite{huang2024revisitingreliabilitypsychologicalscales, rutinowski2024selfperception}, the absence of reported test-retest reliability coefficients prevents final conclusions about their temporal stability \cite{bodroza2024personalitytestingofLLMs}. In contrast, work by Bodroža et al. finds limited temporal stability in LLMs' personality assessments, raising concerns about the trait-like interpretation \cite{ bodroza2024personalitytestingofLLMs}. Similarly, Romero et al. discuss whether personality measurements in LLMs should be considered temporary states or stable traits \cite{romero2023llmssplitpersonality}. 
In total, the current literature provides insufficient and inconclusive evidence to support the claim that LLM personality measurements are temporally stable, calling for further research on this topic. 

\paragraph{3. Consistency Over Situations.}
Personality traits are expected to be consistent in various situations, even if their expression can change. Conceptually, LLMs' fixed architectures could produce a "core" pattern of responses that remains stable in a variety of contexts. Yet, evidence for such cross-situational consistency in LLMs is minimal. 
Several studies demonstrate that LLM responses to personality tests are highly sensitive to prompt variations and contextual cues, leading to substantial variability across situations \cite{huang2024revisitingreliabilitypsychologicalscales, huang2025designingaiagentspersonalitiespsychometric, romero2023llmssplitpersonality, tosato2025persistentinstabilityllmspersonality, shu2024dontneedapersonalitytest, salecha2024LLMsdisplaysocialdesirabilitybias, xie2025aipsychobenchunderstandingpsychometricdifferences, lee2025llmsdistinctconsistentpersonality, frisch2024LLMagentsininteraction, zou2025llmselfreportevaluatingvalidity, huang2025selfreportsmultiobserveragentspersonality, kovac2023llmssuperpositionsculturalperspective}.
Additionally, extensive research has shown that LLMs adopt and tailor their answers to explicitly prompted personality traits \cite{zheng2025LMLPAlinguisticpersonalityassessment, li2025big5chat, jiang2024personallminvestigatingabilitylarge, zeng2025persllmpersonifiedtrainingapproach, molchanova2025exploringpotentiallargelanguage, lotfi2024personalitychatconversationdistillationpersonalized, huang2024revisitingreliabilitypsychologicalscales, jiang2023evaluatinginducingpersonalitypretrained, frisch2024LLMagentsininteraction, li2025evaluatingLLMswithpsychometrics, huang2025designingaiagentspersonalitiespsychometric, tosato2025persistentinstabilityllmspersonality, zou2025llmselfreportevaluatingvalidity, yang2025psyplaypersonalityinfusedroleplayingconversational, bai2025scalinglawllmsimulated, xie2025aipsychobenchunderstandingpsychometricdifferences, lee2025llmsdistinctconsistentpersonality, serapiogarcia2025personalitytraitslargelanguage}. 
Finally, LLMs align their responses to users throughout interactions, resulting in a reduced consistency in their personality profiles \cite{frisch2024LLMagentsininteraction, yang2025psyplaypersonalityinfusedroleplayingconversational, xing2025chameleon}.  
Overall, LLMs' personality scores change due to differences in contextual factors, such as variations in prompts, personality assignments, or conversational partners, undermining their interpretability as context-invariant traits.

\paragraph{4. Inter-Individual Differences.}
The differential perspective emphasizes that personality captures inter-individual differences in behavior and cognition.
Applied to LLMs, this would mean that different LLMs would show distinct Big Five profiles. These differences could be created by variations in training data, learning procedure, model architecture, scale or version.
At the same time, the widespread use of shared safety constraints and system instructions may instead induce uniform behavioral tendencies across models, resulting in common personality profiles.

Empirical findings provide limited support for inter-model differences. Although some studies report variation in personality profiles \cite{bodroza2024personalitytestingofLLMs, sorokovikova2024llmssimulatebigpersonality}, the majority of work documents consistent patterns of a social desirability bias across models \cite{salecha2024LLMsdisplaysocialdesirabilitybias}. Specifically, LLMs tend to score high on Openness for Experience \cite{zheng2025LMLPAlinguisticpersonalityassessment, perez2022discoveringlanguagemodelbehaviors, hilliard2024elicitingpersonalitytraitslarge, tosato2025persistentinstabilityllmspersonality, hartley2025personalitytraitsrisktaking}, Conscientiousness \cite{zheng2025LMLPAlinguisticpersonalityassessment, lee2025llmsdistinctconsistentpersonality, pellert2024AIpsychometrics, tosato2025persistentinstabilityllmspersonality, hartley2025personalitytraitsrisktaking}, and Agreeableness \cite{lee2025llmsdistinctconsistentpersonality, zheng2025LMLPAlinguisticpersonalityassessment, molchanova2025exploringpotentiallargelanguage, pellert2024AIpsychometrics, tosato2025persistentinstabilityllmspersonality, bai2025scalinglawllmsimulated, hartley2025personalitytraitsrisktaking}, and low on Neuroticism \cite{zheng2025LMLPAlinguisticpersonalityassessment, molchanova2025exploringpotentiallargelanguage, pellert2024AIpsychometrics, tosato2025persistentinstabilityllmspersonality, bai2025scalinglawllmsimulated, hartley2025personalitytraitsrisktaking}. Findings for Extraversion are comparatively mixed.
Thus, while it is theoretically thinkable that different LLMs show different response patterns with respect to the Big Five, there is clear evidence for a general pattern of social desirability. This strongly supports the claim there is a universal pattern of the Big Five in LLMs rather than inter-model differences. 

\paragraph{5. Behavioral Relevance.}
Personality traits are latent characteristics that are not directly observable, but they directly influence and cause behavior. Although LLMs lack human-like behavior, it is theorized that their responses to personality tests are relevant to other outcomes, such as safety, toxicity, and persuasion. However, evidence for the predictive validity of LLMs' personality measurements is inconclusive. Several studies report that prompt-induced personality influenced LLM behavior in downstream tasks, such as writing social media posts \cite{serapiogarcia2025personalitytraitslargelanguage} or essays \cite{jiang2023evaluatinginducingpersonalitypretrained}, moral or risk-taking behaviors \cite{huang2025designingaiagentspersonalitiespsychometric, hartley2025personalitytraitsrisktaking} and the agent's language \cite{frisch2024LLMagentsininteraction}. While these findings support behavioral relevance of LLM personality measurements, other studies found that LLMs' self-reported personality traits do not consistently predict aspects of their behavior, including social-bias, honesty, sycophancy or interaction quality \cite{ai2024selfknowledgeactionconsistentnot, zou2025llmselfreportevaluatingvalidity, han2025personalityillusionrevealingdissociation}. In sum, there is mixed evidence that LLMs' responses to personality tests are relevant for downstream behavior.

\paragraph{6. Descriptive Summaries.}
The Big Five were developed through a lexical approach, summarizing words that characterize human personality traits \cite{larsen2017personalitypsychology, matthews2003personalitytraits}. It is highly implausible that traits created for humans are similarly meaningful to describe LLMs. Using human-like words to describe LLMs contributes to the anthropomorphization of these models. Items from established personality questionnaires, such as "I am the life of the party" or "I leave my belongings around" \cite{goldberg1992bigfive}, are not applicable to LLMs. 
This limitation is strongly supported by  evidence in the literature. Researchers found that LLMs frequently refuse to answer personality test items \cite{lee2025llmsdistinctconsistentpersonality, xie2025aipsychobenchunderstandingpsychometricdifferences, shu2024dontneedapersonalitytest}. Sühr et al. argue that personality tests are built for humans, not LLMs, and found that they are not measurement invariant between humans and LLMs, \emph{i.e.,} their responses structurally deviate \cite{suehr2024challengingvaliditypersonalitytests}. Additionally, scholars have recently developed new personality taxonomies for AI agents \cite{voelkel2020developingapersonalitymodelpsycholexical, kovacevic2024personalitydimensionsgpt}, as the "Big Five model for human personality does not adequately describe agent personality" \cite{voelkel2020developingapersonalitymodelpsycholexical}. In short, human personality traits do not provide a useful description of LLMs, neither in theory nor in practice.

\section{Research Agenda: Rethinking Personality for LLMs}
The findings show that personality tests do not measure a construct in LLMs equivalent to human personality, as LLMs satisfy none of the six defining characteristics sufficiently.
This calls for rethinking what these measurements actually capture. We outline a research agenda that moves beyond anthropomorphic assumptions and distinguishes three concepts: functional LLM behavior, malleable interaction-relevant patterns, and stable intrinsic characteristics of LLMs.

\paragraph{Personality Tests Do Not Measure Traits in LLMs.}
A primary challenge lies in the uncritical transfer of human personality frameworks to artificial systems. 
Personality inventories do not measure stable traits in LLMs comparable to human personality.
Rather than labeling LLMs with human traits, research should move toward functional evaluations and task-specific behavior.
Instead of asking whether an LLM is “agreeable” or “conscientious”, researchers might assess how different response styles affect outcomes in human-AI collaboration. 
Alternative frameworks could focus on contextual robustness, interaction style, or prompt sensitivity. These dimensions could better capture LLMs' functional behavior across tasks and situations \cite{karra2023estimatingpersonalitywhitebox, li2025evaluatingLLMswithpsychometrics}. 
Characterizing LLMs along these dimensions shifts the focus from anthropomorphic labels toward evaluating LLM behavior in specific contexts.

\paragraph{What Personality Tests Actually Capture.}
While personality tests do not measure stable traits in LLMs, they might capture malleable behavioral patterns relevant for human-agent collaborations. 
Since users often treat LLMs as social interaction partners, personality tests might serve as a practical tool for characterizing these interactions. They can provide standardized vocabulary without assuming that LLMs possess genuine human-like personality traits. 
However, the validity of these instruments for LLMs must be empirically tested, not assumed. 
Future work should clarify what these measurements actually capture and explicitly distinguish between malleable personality-like response patterns and stable characteristics in LLMs.

\paragraph{Stable Characteristics in LLMs.}
LLMs may possess stable intrinsic characteristics that are different from human personality but fulfill the six defining criteria of a personality trait. Such characteristics would enable comparing LLMs based on their inherent, stable behavioral tendencies.
Future research should identify which properties meet the six characteristics of a trait. Potential candidates include conversational alignment, instruction adherence strictness, stylistic tendencies, and ability to adapt. 
Researchers should validate any instruments used to measure LLM traits, testing for test–retest reliability, cross-situational consistency, and measurement invariance.

\paragraph{Bias, Normativity and Social Impact.}
LLMs display socially desirable traits, raising concerns about normative bias and implicit cultural expectations.
Future work should examine whose values are reflected in these LLM profiles and how they impact users. This includes whether certain personas affect user trust and reliance, influence vulnerable users or manipulate users in high-stakes decision-making contexts.

\section{Conclusion}
This paper questions the application of human personality tests to LLMs, as none of the six defining characteristics of personality are fully fulfilled when applied to LLMs. 
However, such psychological instruments may capture meaningful variation in LLM outputs relevant for human-AI collaboration. Rather than anthropomorphizing LLMs, personality measures can serve as a metaphor for interactional patterns users value in both human-human and human-agent relationships.
We propose three research directions: shift toward functional evaluation of LLM behavior, clarify what personality tests actually measure in LLMs, and identify stable intrinsic characteristics that could serve as an LLM-specific trait framework.

\begin{acks}
This work has been partially supported by a nominal grant received at the ELLIS Unit Alicante Foundation from the Regional Government of Valencia in Spain (Resolución de la Conselleria de Industria, Turismo Innovación, y Comercio, Dirección General de Innovación) and by Intel. K.Z. has also been partially funded by the Bank Sabadell Foundation. 
\end{acks}


\end{document}